# Multi-Objective Optimization for Sustainable Closed-Loop Supply Chain Network Under Demand Uncertainty: A Genetic Algorithm

Ahmad Sobhan Abir
Dept. of Industrial and
Production Engineering
Rajshahi University of
Engineering & Technology
Rajshahi-6204; Bangladesh
abir145011@gmail.com

Ishtiaq Ahmed Bhuiyan
Dept. of Industrial and
Production Engineering
Rajshahi University of
Engineering & Technology
Rajshahi-6204; Bangladesh
ishtiaq145022@gmail.com

Mohammad Arani
Dept. of Systems Engineering
University of Arkansas at Little
Rock
2801 S. University Ave., Little
Rock, AR 72204, USA
mxarani@ualr.edu

Md Mashum Billal
Dept. of Mechanical Engineering
(Engineering Management)
University of Alberta, Edmonton
116 St & 85 Ave, Edmonton, AB
T6G 2R3, Canada
mdmashum@ualberta.ca

*Abstract*—Supply chain management has been concentrated on productive ways to manage flows through a sophisticated vendor, manufacturer, and consumer networks for decades. Recently, energy and material rates have been greatly consumed to improve the sector, making sustainable development the core problem for advanced and developing countries. A new approach of supply chain management is proposed to maintain the economy along with the environment issue for the design of supply chain as well as the highest reliability in the planning horizon to fulfill customers' demand as much as possible. This paper aims to optimize a new sustainable closed-loop supply chain network to maintain the financial along with the environmental factor to minimize the negative effect on the environment and maximize the average total number of products dispatched to customers to enhance reliability. The situation has been considered under demand uncertainty with warehouse reliability. This approach has been suggested the multi-objective mathematical model minimizing the total costs and total CO2 emissions and maximize the reliability in handling for establishing the closed-loop supply chain. Two optimization methods are used namely Multi-Objective Genetic Algorithm Optimization Method and Weighted Sum Method. Two results have shown the optimality of this approach. This paper also showed the optimal point using Pareto front for clear identification of optima. The results are approved to verify the efficiency of the model and the methods to maintain the financial, environmental, and reliability issues.

I. INTRODUCTION

One of the most uprising topics in the current decade has been the sustainable supply chain, as the environment is polluted at an accelerating rate owing to corporate and factory functioning. Due to the carbon and the product produced by factories, environmental degradation is growing day by day. The use of the green supply chain is usually caused by two of the most prevalent interrelated variables, i.e. the number of reducing emissions and the price of implying a sustainable supply chain [1]. Environmental issues have already been concerned in developed countries and government law and customer pressure have increased to follow SSC and considering environmentally friendly production. Supply chain design models have generally worked on minimizing costs without considering carbon emissions. But in recent times so many studies like [2]–[6] focused on environmentally friendly production thus taking carbon emission into account and optimization total cost. As a result, industries should now have been compelled to redesign their supply chain network to deal with green production growth. As a consequence, SSC has supported comprehensive decision-making to implement economic issues with environmental perception. Besides, if a distribution layer plant has been unable to service the lowest levels of the chain for reasons such as natural occurrences, acts of terrorism, change of owners, labor mistakes, weather conditions, etc. Reallocation of clients to other active distributors shifts the supply chain topology and thus considerably reduces expenses. It is therefore now important to develop a system that will decrease both CO2 emissions and total costs and also improve reliability to maximize the average total number of products dispatched from warehouses to consumers.

In our research, we have run a mathematical model to optimize both the carbon emission and the total cost. Here, multi-objective optimization has been identified using two mathematical approaches both the exact solution method and meta-heuristic approach. For the green approach, the minimization of carbon emission as well as minimization of the total cost has been identified. A closed-loop supply chain network has also been incorporated with reverse flow. This closed-loop mainly deals with the end life of products and unused raw materials that otherwise it can create more wastage and degradation of the environment.

A literature review has been provided to find out the previous work and thus the specified characteristics of our contribution to the body of knowledge. In that context, we have also evaluated the problem statements, and the mathematical gap, as well. In the sustainable supply chain model optimization, most of the authors considered a single objective or open-loop supply chain or merely just the production. Though some works have evaluated multi-objective optimization, from the best of our knowledge, the new problem statement accompanying solution methodology and an analyzing comparison is the novelty of this paper. This motivated us for performing this research. We also showed the relationship between the solution sets of two methodologies mentioned earlier, the epsilon constraint method, and the genetic algorithm optimization.

II. LITERATURE REVIEW

Given their significance in the contemporary economy, supply chains have for many years been on the study agenda of a multitude of companies and other academic disciplines. Reference [7] proposed a carbon emission limitation supply chain model. Their model of optimization for sustainable supply chain management demonstrated the relationship between environmental management and its impact on the supply chain. They used a mixed-integer (MIP) program to optimize the organization's carbon emissions while minimizing the cost of opportunity. Reference [8] concentrated on a computer and electronic components of an organization's GSCM practice. Their work was primarily based on raising consideration of environmentally friendly supply chain management. They focused on product manufacturing, procurement choices, and reverse logistics from their studies.

Reference [9] constructed a supply chain network model concerning environmental issues. The concern of environmental factors at the design stage and proposed a multi-objective optimization model to enhance investment on environmental factors. This research focused on the total cost of environmental influence on the level of available environmental facility and also influence on handling and transportation activities. Reference [10] provided Operations Research's effect and contribution to green logistics, which is an integration of environmental elements. Taking into consideration the design, planning, and monitoring of the supply chain with transport, inventory, facilities issue, etc., they suggested a model that described the possible trends. Reference [11] worked to identify and analyze different research to find out the definitions and interrelationship between sustainable supply chain and sustainable supply chain management. They focused on the two key characteristics of business sustainability and supply chain management to analyze the definitions and evaluate the interrelationship.

Reference [12] worked to reduce carbon emissions in the general environmental supply chain network. They also looked at the complete supply chain price. To explore this problem, they used solid fuzzy programming. To demonstrate and solve this issue of optimization, they regarded the π-restriction approach and numerical model. Reference [13] explored traditional supply chain management and demonstrated that the supply chain should be more robust and effective with growing levels of industrialization and globalization. They concentrated on sustainable development and suggested a fresh supply chain network that considered environmental and economic problems. To minimize expenses and carbon emissions, they regarded a multi-objective optimization model.

Reference [14] explored two significant variables in sustainable development production planning. They regarded these two goals in the parallel account of carbon emission and energy consumption and suggested production planning through current and new process paths to minimize one goal while maximizing the other goal. They suggested a graphical method based on pinch analysis ideas. Most authors interpreted a single goal or open-loop supply chain, or just development, in the optimization of the sustainable supply chain model. While some works have assessed multi-objective optimization, no work uses both heuristic and direct solutions to two unique mathematical models. This prompted us to do this study. We also showed the relationship between these two mathematical models for clear identification and more acceptability.

Reference [15] examined how GSC influenced the company's regulatory, market, and competitive decisions and strategy in China. Mainly the research was based on the overall effect of manufacturing products on ecological and environmental factors. Here the organizations are bound to take the initiative of GSCM in response to customer pressure and to retain market competitiveness. The main problem was they didn't fully examine the cause and number of isomorphic institutional pressures. Also, the firm size was not considered in response to the pressure. References [16], [17] constructed a model for a manufacturing firm of southern India consists of the drivers comprising the execution of GSC employing an Interpretive Structural Modeling appearance. The different numerous drivers of GSC are ascertained emerged on the GSM counsels with specialists in the industry. The model has the limitations same as the Interpretive Structural Modeling methodology. The model used in this research is extremely dependent on the adjudication of the specialist team which is a human factor with its limitations.

Reference [18] constructed a framework to exhibit a group of indices for appraising the representation of the vehicle GSC. This study investigated various literature on GSC concerning measurement, environmental management, strategic supply chain evaluation, and automobile supply chain management. To extensively and efficiently set up the pertinent measures, a competent model concerning the vehicle GSC as a two-in-one chain was provided. The prime limitation of this study is that the survey was conducted based on expert evaluation not concerning industrial actual evaluation and experiences. Reference [17] presented an approach for GSC considering carbon trading and green procurement as environmental factors. The manufacturer has taken environmental consideration into account owing to the increasing pressure because of climate change. The negative effect on the environment owing to the organization's work was considered. They formulated a mixed-integer program that minimized the total costs of the traditional GSC.

Reference [19] investigated total factor $CO_2$ emission performance (TFCEP) that effects on green development. The global Malmquist-Luenberger index method was used to evaluate the total carbon emission or TFCEP. They examined the determinants of $CO_2$ emission estimation with the help of a generalized method. They formulated a practical method to boost the TFCEP index. To enhance the TECEP, they approached to motivate green technological innovation and showed that adapting this technology affects the environment satisfactorily. Reference [20] proposed a two-stage stochastic programming model for the GSC network. The model considered the location problem and carbon trading. This study integrated previous literature with carbon pricing and product demand. This model can also be incorporated with the uncertainty of carbon price and product demand.

Pollution destroys the natural balance of the environment. Day by day the rate of pollution is alarmingly increasing. Pollution runs parallel to civilization. Although, it started slowly, as time elapsed the pollution and its adverse effect have deteriorated. However, this topic remained unnoticed up to the 20th century. As a result, several agreements have been signed on compliance with environmental objectives. Bangladesh's government has also taken some measures to protect the environment. Two most known steps are:

- Environment Policy, 1992.
- Environment Conservation Rules, 1997.

So, environmental pollution must be controlled for our existence maintenance. Organizations must take steps thus to follow the rules and regulations to ensure that environment is not affected by their conduct. So, greening is a must to save the planet, and organizations must consider the environment-friendly production system.

## III. PROBLEM STATEMENT

Global warming and pollution of the environment have become a major problem. Every time, industries emit massive amounts of greenhouse gases. $CO_2$ is the primary component of these gases. We have nothing to do to get rid of the future catastrophe but to reducing the rate of $CO_2$ emissions from those industries. The developed countries are introducing $CO_2$ emission taxes. To this end, supply chain costs have risen. Besides, if a distribution layer plant has been unable to service the lowest levels of the chain for reasons such as natural occurrences, acts of terrorism, change of owners, labor mistakes, weather conditions, etc. Reallocation of clients to other active distributors shifts the supply chain topology and thus considerably reduces expenses. It is therefore now important to develop a system that will decrease both $CO_2$ emissions and total costs and also improve reliability to maximize the average total number of products dispatched from warehouses to consumers. Fig. 1 illustrates the closed-loop supply chain in which the disposal activity is one of the network's responsibilities.

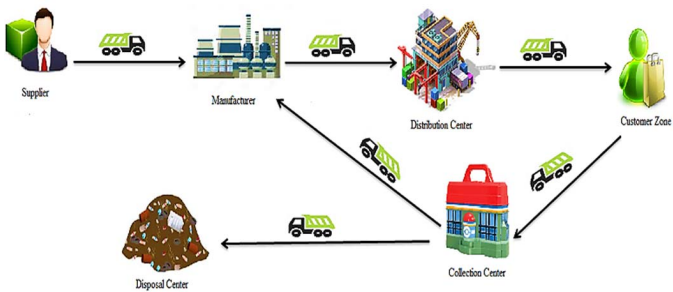

Fig. 1. Example of Closed-Loop Supply Chain Network [21].

The GSC problem mathematical model consists of six components, namely supplier, manufacturers, distribution centers, customers, customer center, and disposal center which are shown in Fig. 1. It is assumed to be one product solely. The composition of the network is not necessarily the same, since a specified quantity of commodity units is assigned by these customers and specific arrangements are made at each point. Three types of products are formed (new product, product for disposal and product for demolition) through the mathematical model through the supply chain network, in which the product description, based on the product structure, is as follows:

- A new product, whether new or manufactured, would be forwarded along the network line (manufacturer to a distribution center and distribution center to the customer) to meet customer demand.
- The substance to be disposed of is purchased from the consumer and delivered to the disassembly center.
- The product to be demolished is distributed from the disassembly center to the manufacturer.

### A. Objectives

- To minimize the total supply chain cost (the total supply chain cost denotes the total fixed cost, the total variable cost, and the total transportation cost).
- To minimize the total $CO_2$ emissions (total $CO_2$ emissions denotes the $CO_2$ emissions of production, assembly, handling, disassembly, remanufacturing, and transportation).
- To maximize reliable dispatching of product from warehouse to customer transportation.

### B. Formulation

For our research paper which is to propose a new multi-objective Genetic Algorithm as a new solution methodology, we employed a mathematical framework proposed by [22], [23] and its benchmark data.

## IV. GENETIC ALGORITHM

The genetic algorithm is based on the principles of natural selection and evolutionary genetics. They combine the survival of the most powerful string systems. The framework requires a randomized exchange of information to construct a ground-breaking algorithm for optimal searching. Genetic Algorithms developed by Holland (1975), a population-based probabilistic and optimizing technique. They are classified as global search heuristics and are also a class of evolutionary algorithms using techniques triggered by evolutionary biologists such as inheritance, mutation, selection, and crossover (also known as recombination). We are implemented as a candidate solutions computer simulation to the problem of optimization which communicates a better solution. Solutions are conventionally represented in the binary strings 0s and 1s, but any encoding is not impossible. Usually, evolution starts with a population of randomly created individuals and happens in generations. The fitness of each person in the population is calculated in each generation, multiple individuals are selected from the current population (based on fitness), and modified (recombined and potentially mutated) to create a new population. Even the new population is being used for the next algorithm iteration. Normally, the algorithm is completed when either a maximum number of generations has been generated or an optimal fitness level for the population has been reached. If the algorithm is terminated due to a maximum number of generations, there may or may not have been an optimal solution. For further information, refer to [24], [25]. Fig. 2 depicts the Pareto Frontier of the multi-objective genetic algorithm, moreover, as an example of the solution structure, TABLE I. provides a list of decision variables of our concern.

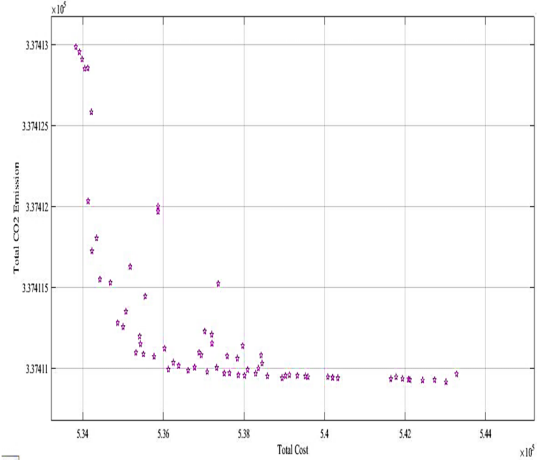

Fig. 2. Pareto Frontier Provided by GA.

*A. Working Cycle of a GA*

1. The initial population of individuals is created without discrimination.
2. The fitness of every person in the population is measured.
3. The evaluation function is given by the objective function and it gives a score to any individual.
4. Two individuals are selected based on their fitness, the greater the fitness, the higher the probability of being selected.
5. These individuals then "reproduce" to create one or more offspring, after which the offspring are randomly mutated.
6. This procedure continues until an optimal solution has been discovered or a certain number of generations have been passed, depending on the accuracy we need.

*1) Initialization*

Originally, individual solutions are created without prejudice to shape a primary random population. The size of the population depends on the characteristics of the problem, but it symbolically holds several hundred or thousands of possible solutions. Traditionally, the population is produced without discrimination, covering a wide range of possible solutions (within the search space).

*2) Selection*

During each generation, a proportion of the current population is chosen to replicate the new generation. Individual solutions are chosen through a fitness-based process, where fitter solutions (as measured by fitness functions) are commonly chosen with a higher chance. Selection methods measure the fitness of each solution and make the most of the best solutions. Another approach only scores a random sample of the population, as the procedure can be time-consuming. Most selection functions are stochastic such that the proportion of less perfect solutions has a lesser chance of being chosen.

*3) Crossover*

Crossover is a mechanism that produces new individuals by joining parts from two Individuals. Crossover is experimental which makes a huge jump in between two areas. Single point, Multipoint, and identical crossovers are available. Simulated Binary Crossover produces children that differ from their parents.

*4) Mutation*

A mutation is a mechanism that produces new individual(s) by creating alternates in a single chromosome rather than the entire string. The mutation is experimental that produces indiscriminate small deviations, so staying close to the parent. Simply mutation can present new information.

*5) Termination*

This process is repeated until the termination condition has been succeeded. Common terminating conditions are:

1. The optimal solution which meets the minimum criteria is obtained.
2. Followed a predefined number of generations.
3. Exhausted computational resources.
4. The fitness of the highest-ranking solution has reached a significant proportion that potential iterations no longer produce better results.
5. Manual inspection of solution guides no further search.
6. Any of the above combinations.

TABLE I. ONE POINT FROM PARETO FRONTIER

| Variables | Values | Variables | Values |
|---|---|---|---|
| $Xa_1$ | 0.719 | $Xa_2$ | 0.673 |
| $Xb_1$ | 0.805 | $Xb_2$ | 1.304 |
| $Xd_1$ | 0.724 | $Xd_2$ | 0.758 |
| $Ya_{11}^1$ | 1629.999(units) | $Ya_{12}^1$ | 1629.999(units) |
| $Ya_{21}^2$ | 949.999(units) | $Ya_{22}^2$ | 949.999(units) |
| $Yb_{11}^1$ | 1629.999(units) | $Yb_{12}^1$ | 949.999(units) |
| $Yb_{21}^2$ | 1629.999(units) | $Yb_{22}^2$ | 949.999(units) |
| $Yc_{11}^1$ | 325.999(units) | $Yc_{12}^1$ | 325.999(units) |
| $Yc_{21}^2$ | 142.499(units) | $Yc_{22}^2$ | 142.499(units) |
| $Yd_{11}^1$ | 32.599(units) | $Yd_{12}^1$ | 14.249(units) |
| $Yd_{21}^2$ | 16.299(units) | $Yd_{22}^2$ | 7.125(units) |

V. CONCLUSIONS

Because of the environmental issue, the sustainable supply chain region is currently increasing. A business primarily believes in economic reality and environmental concerns as well. So, with the multi-objective approach, this paper suggests a mathematical model of the sustainable supply chain and optimizes the sustainable supply chain network. Total expense and complete $CO_2$ emissions are the two primary criteria as well as maximizing reliability to assess which are chosen to create a linear mixed-integer model. A Pareto front represents the three objectives of the total cost, the total emission of $CO_2$ and warehouse reliability. One close-loop network design can be achieved by the calculation of these three objectives. The assumptions construct the model more realistic. This model includes only one transportation system for calculation. The suggested Genetic Algorithm Optimization attempts to minimize the total cost along with total emission of CO2 and enhancing reliability. Over the last few years, researchers have suggested a number of models and reports for sustainable supply chain design with varying degrees of sophistication. This study is an effort to join the bridge of reliability, the total cost of production, and total CO2 emission as well as finding the ideal point for these three goals. The suggested strategy is therefore expected to be appropriate for making sustainable supply chain choices in the actual globe. For future research directions, further analysis can then be proceeded by constructing a new model; other optimization techniques such as Discrete Event Simulation (DES) [26] can be used to tackle the uncertainty, vehicle routing problem could be an integrated part of the research [27], [28], additionally, for the delivery of products, different modes of transport and vehicle maintenance can be considered [29].

ACKNOWLEDGMENT

The authors would like to express their gratitude and sincere appreciation to the authors of [22], [23] for allowing us to exploit the potentials of their proposed mathematical model as a vehicle for serving the purposes of this research.